\documentclass[10pt,draftcls,onecolumn,journal]{IEEEtran}
\usepackage{amssymb}
\usepackage{latexsym}
\usepackage{graphicx}
\usepackage{psfig}

\begin{document}

\newcommand{\qed}{\rule{7pt}{7pt}}
\newcommand{\calc}{{\cal C}}
\newcommand{\F}{{\mathbb F}}
\renewcommand{\vec}[1]{{\mathbf #1}}
\newcommand{\ip}[1]{{\langle #1 \rangle}}
\newcommand{\tup}[1]{{\left( #1 \right)}}
\newcommand{\FOR}{{\bf for}}
\newcommand{\mytab}{\hspace*{1.5em}}
\newcommand{\sem}[1]{{\sf{\em #1}}}
\newcommand{\Z}{{\mathbb Z}}
\newtheorem{theorem}{Theorem}[section]
\newtheorem{lemma}{Lemma}[section]
\newtheorem{corollary}{Corollary}[section]
\newtheorem{observation}{Observation}[section]
\newtheorem{proposition}{Proposition}[section]
\newtheorem{definition}{Definition}[section]
\newtheorem{claim}{Claim}[section]
\newtheorem{fact}{Fact}[section]
\newtheorem{assumption}{Assumption}[section]
\newtheorem{example}{Example}[section]
\newtheorem{property}{Property}[section]
\newtheorem{assertion}{Assertion}[section]
\newcommand{\R}{{\mathbb R}}
\newcommand{\kwd}[1]{\ {\bf \small #1}\ }

\title{Efficient Convergent Maximum Likelihood Decoding on Tail-Biting Trellises\footnote{The results in this paper appear in part in ISIT 2001\cite{sh2}}}
\author{ \authorblockN{Priti Shankar\thanks{Priti Shankar acknowledges
      support fron the Scientific Analysis Group, DRDO, Delhi}\authorrefmark{1}, P.N.A.Kumar\authorrefmark{2}, K.Sasidharan\authorrefmark{3}, B.S.Rajan\authorrefmark{4}, A.S. Madhu\authorrefmark{1}\\}
\authorblockA{\authorrefmark{1}
Department of Computer Science and Automation, Indian Institute of Science, Bangalore, India\\ Email: \{priti,madhu\}@csa.iisc.ernet.in\\}
\authorblockA{\authorrefmark{2} Microsoft Silicon Valley Campus,
  Mountain View, California.\\  
Email: pnakumar@yahoo.com\\ }
\authorblockA{\authorrefmark{3} Veritas Software India, Bund Garden Road, Pune, India\\Email: sasi@veritas.com\\}
\authorblockA{\authorrefmark{4}Department of Electrical Communication
  Engineering, Indian Institute of Science, Bangalore, India\\ Email: bsrajan@ece.iisc.ernet.in}}

\maketitle

\begin{abstract}
An algorithm for exact maximum likelihood(ML) decoding on tail-biting
trellises is presented, which exhibits very good average case
behavior. An approximate variant is proposed, whose simulated
performance is observed to be virtually indistinguishable from the
exact one at all values of signal to noise ratio, and which
effectively performs computations equivalent to at most two rounds on
the tail-biting trellis. The approximate algorithm is analyzed, and
the conditions under which its output is different from the ML output
are deduced. The results of simulations on an AWGN channel for the
exact and approximate algorithms on the 16 state tail-biting trellis
for the (24,12) Extended Golay Code, and tail-biting trellises for two
rate 1/2 convolutional codes with memories of 4 and 6 respectively, are reported. An advantage of our algorithms is that they do not suffer from the effects of limit cycles or the presence of pseudocodewords.
\end{abstract}

\section{Introduction}
\label{sec:intro}
Tail-biting trellises are perhaps the simplest instances of decoding
graphs with cycles. A tail-biting trellis has a Tanner
graph~\cite{tanner} with a single cycle and usually approximate
algorithms are used for decoding, as exact algorithms are believed to
be too expensive. These approximate algorithms iterate around the
trellis until either convergence is reached, or for a preset number of
cycles. To the best of our knowledge, no {\em exact} decoding
algorithms other than the brute force algorithm have been proposed so
far for the general case, though there are several {\em approximate}
algorithms for maximum-likelihood decoding~\cite{shu,ma,
  chep,wang,sundberg,ku} and exact algorithms for bounded distance
decoding~\cite{cav}. The problem of Maximum A-Posteriori
Probability(MAP) decoding is not addressed here. We propose an {\em
  exact} recursive algorithm, which exhibits very good average case
behavior.  The algorithm exploits the fact that a linear tail-biting
trellis can be viewed as a coset decomposition of the group
corresponding to the linear code with respect to a specific subgroup
and is based on the $A^*$ algorithm~\cite{nil}. We also propose two approximate
variants that always converge, and observe their performance on
tail-biting trellises for the (24,12) extended Golay code and two
convolutional codes of rate 1/2 and memory of 4 and 6
respectively. The performance of the first approximate variant is
indistinguishable from that of the exact algorithm  in terms of bit
error rate for the two convolutional codes, and it is guaranteed to
update each node in the tail-biting trellis at most twice i.e it
performs a computation equivalent to at most two rounds on the
trellis. Section ~\ref{related} briefly mentions related
work. Section~\ref{sec:background} provides some
background. Section~\ref{algo} describes the algorithm, while
Section~\ref{analysis} analyses the algorithm. Section ~\ref{approx}
describes the approximate algorithm and provides an analysis for its
good performance.  Section~\ref{simulations} reports the results of simulations on an AWGN channel and section~\ref{conclusion} concludes the paper.
\section{Related Work}
\label{related}
Aji et al.~\cite{aji} have shown that iterative maximum-likelihood (ML)
decoding on tail-biting trellises will asymptotically converge to
exact maximum likelihood decoding for certain codes. They provide
experimental evidence that practically ML decoding is achieved for the
$(8,4)$ Hamming code with five rounds of the tail-biting trellis. The
presence of {\em pseudocodewords} sometimes results in sub-optimal
decoding and it is also possible to have situations where the
iterative message passing algorithm does not converge.  Several
maximum likelihood decoding algorithms on tail-biting trellises have
been proposed without a theoretical
analysis~\cite{ma,wang,chep,solomon,shu,ku}, but with good experimental
results. Most of these are sub-optimal algorithms in that they may not
produce the exact maximum-likelihood result on termination. Anderson
and Hladik~\cite{cav} have given an algorithm that is optimal for
bounded distance decoding. The $A^*$ algorithm~\cite{nil} has been used for
maximum likelihood soft decision decoding on conventional trellises for block codes
~\cite{han1,ekroot,ksch98,han2,han3,shih}. In ~\cite{han1}
Han et al. propose the use of the $A^*$ algorithm for ML decoding of block
codes on their conventional trellises and report significant experimental gains in
decoding complexity for signal to noise ratios ranging from 5 dB to 8
dB. This algorithm has been analyzed in \cite{han5} and shown to be
efficient for many practical communication systems.  In
\cite{han2} a modified algorithm is proposed which searches through
error patterns instead of codewords and similar
gains are reported. Heuristic search algorithms are proposed in \cite{shih} which
combine previously proposed algorithms and are able to outperform other
practical decoders. A tutorial paper on the application of the $A^*$
algorithm to soft decision decoding appears in \cite{ekroot}. Sorokine and Kschischang~\cite{ksch98} propose a metric called
the variable bias term that is used in an $A^*$ algorithm, which has
low computational complexity. Aguado and
Farrell \cite{aguado} discuss modified sequential algorithms on
conventional trellises for block
codes, which offer reduced
complexity in comparison with the original stack algorithm
\cite{jelinek} for sequential decoding. Han et
al. \cite{han4} propose a trellis based ML soft-decision decoder for
convolutional codes which uses a
stack and a metric that ensures ML decoding.
\section{Background}
\label{sec:background}
We first present some background on tail-biting trellises.
Tail-biting trellises for convolutional codes were introduced in
\cite{solomon}. Minimal tail-biting trellises for block codes have been discussed in~\cite{cald,kv1,kv2}. 
\begin{definition}
A {\em tail-biting trellis} $T = (V, E, \F_q)$ of depth $n$ is an edge-labeled
directed graph with the property that the set $V$ can be partitioned
into $n$ vertex classes
\begin{equation}
\label{eq:parti}
V = V_0 \cup V_1 \cup \ldots \cup V_{n-1}
\end{equation}
such that every edge in $T$ is labeled with a symbol from the alphabet
$\F_q$, and begins at a vertex of $V_i$ and ends at a
vertex of $V_{i+1(mod \ n)}$, for some $i \in \{0,1, \ldots , n-1\}$. 
\end{definition}
 We identify ${\cal I}$ the set of time indices 
with $\Z_n$, the residue classes of integers modulo $n$. An interval of indices
$[i,j]$ represents the sequence $\{i,i+1,\ldots j\}$ if $i<j$, and the sequence
$\{i,i+1,\ldots n-1,0,\ldots j\}$ if $i>j$. Every cycle in $T$ starting at a
vertex of $V_0$ defines a vector 
$(a_1, a_2, \ldots, a_n) \in \F_q^n$ which is an 
{\em edge-label sequence}. We assume that every vertex and every edge in the 
tail-biting trellis lies on some cycle, that is the tail-biting
trellises we are dealing with are {\em reduced} \cite{kv2}. The trellis $T$ represents a block
code $\calc$ over $\F_q$ if the set of all edge-label sequences in $T$ is
equal to $\calc$. Let $\calc(T)$ denote the code represented by a trellis $T$.

A linear tail-biting trellis, 
 for an $(n,k)$ linear block code
$\calc$ over $\F_q$ can be constructed as a {\em trellis product}~\cite{ksch95} of the 
representation of the individual trellises (called {\em elementary }
trellises) corresponding to each of the $k$ rows of the 
generator matrix $G$ for $\calc$~\cite{kv2}. Let $T_1$ and $T_2$ be
the component trellises. The set of vertices $V_i(T_1 \times T_2)$ of the product trellis
$T_1\times T_2$ at time index $i$, is just the Cartesian product of
the vertices of the component trellises. Thus 
$V_i(T_1\times T_2)$= $V_i(T_1)\times V_i(T_2)$. Consider
$E_i(T_1)\times E_i(T_2)$, and interpret an element
$((v_1,\alpha_1,v_1'),(v_2,\alpha_2,v_2'))$ in this product, where
$v_i,v_i'$ are vertices and $\alpha_1,\alpha_2 $ edge labels, as the
edge $((v_1,v_2), \alpha_1+ \alpha_2, (v_1',v_2'))$ where $+$ denotes
addition in the field $\F_q$. If we define the $i^{th}$ section as the set of edges connecting the vertices at time index $i$ to those at time index $i+1$, then the edge count in the $i^{th}$ section is the product of the edge counts in the $i^{th}$ section of the individual trellises. 

Let $\left\{\vec{g_1}, \vec{g_2}, \ldots , \vec{g_k}\right\}$ be the rows of a
generator matrix $G$ for the linear code $\calc$.
Each vector $\vec{g_i}$ generates a one-dimensional subcode of $\calc$,
which we denote by $\ip{\vec{g_i}}$. Therefore 
$\calc = \ip{\vec{g_1}}+\ip{\vec{g_2}} + \cdots +\ip{\vec{g_k}}$, and the
trellis representing $\calc$ is given by $T = T_1 \times T_2 \times \cdots
\times T_k$, where $T_i$ is the trellis for $\ip{\vec{g_i}},\ 1 \leq i \leq k$.
To specify the component trellises in the trellis product above,
we will need to introduce the notions of linear\cite{ksch95} and circular spans\cite{kv2} 
and elementary trellises~\cite{ksch95,kv2}. 
Given a codeword $\vec{c}=(c_1,c_2,\ldots c_n) \in \calc$, the {\em linear span} of
$\vec{c}$, is the smallest
interval $[i,j] \in I= \{1,2,\ldots n\}, i\leq j$ which contains all the non-zero positions of $\vec{c}$. 
A {\em circular span} has exactly the same definition with $i>j$. Note that
for a given vector, the linear span is unique, but circular spans are 
not-- they 
depend on the runs of consecutive zeros chosen for the complement of the span 
with respect to the index set $I$. For a vector 
$\vec{x} = \tup{x_1, \ldots , x_n}$ over the field $\F_q$ and a
specified span $[i,j]$, there is a
unique {\em linear elementary trellis} representing $\ip{\vec{x}}$~\cite{kv2}. This 
trellis has $q$ vertices at time indices $i$ to $(j-1)$ mod $n$, and a
single vertex at other positions. 
Consequently, $T_i$ in the trellis product mentioned earlier, is the 
elementary trellis representing $\ip{\vec{g_i}}$ for some choice of 
span (either linear or circular).
Koetter and Vardy~\cite{kv2} have shown that any linear trellis, conventional or
tail-biting can be constructed from a generator matrix whose rows can be
partitioned into two sets, those which have linear span, and those taken to
have circular span. The trellis for the code is formed as a product of the elementary
trellises corresponding to these rows. We will represent such a generator
matrix as 
$G_{KV} = \left[\begin{array}{c} G_l \\ \hline G_c \end{array}\right]$, where 
$G_l$ is the submatrix consisting of rows with linear span, and $G_c$ the 
submatrix of rows with circular span.
\begin{definition}
\label{def:zerorun}
 For a vector $\vec{v}$ of
  circular span $ [i,j]$ in $G_c$, the interval $[j$ mod $n, (i-1)$ mod $ n]$ is
  called the {\em zero run} of the vector.
\end{definition}

 The path in the trellis
  corresponding to this vector shares all states at time indices in the zero run with
  the path corresponding to the all-zero codeword in the product
  trellis.\\
For example,consider the codeword $0100011$ with circular span
  $[6,2]$. This has zero run $[2,5]$. The elementary trellis
  corresponding to this vector has state cardinality profile
  $(2,2,1,1,1,1,2)$. (Recall the time indices are numbered from 0 to
  $n-1$ where $n$ is the length of the code).

As an example we display a tail-biting trellis for a binary $(7,4)$
Hamming code. Though this is not a minimal trellis for the code, it serves to
illustrate  some of the definitions above. The spans of the rows are
shown alongside the rows. All spans $[i,j]$ with $i$ greater than $j$
are circular spans.

\begin{example}
\label{ex:overlay2}
Let $\calc$ be a $\tup{7,4}_2$ Hamming code a with a product generator matrix $G_{KV}$ defined as
\[
G_{KV} = \left[\begin{array}{ccccccc}
1 & 0 & 0 & 0 & 1 & 1 & 0 \\
0 & 0 & 1 & 0 & 1 & 1 & 1 \\
\hline
0 & 1 & 0 & 0 & 0 & 1 & 1 \\
0 & 1 & 1 & 1 & 0 & 0 & 1
\end{array}
\right]
\begin{array}{c}
\lbrack 1,6 \rbrack \\
\lbrack 3,7 \rbrack \\
\lbrack 6,2 \rbrack \\
\lbrack 7,4 \rbrack
\end{array}
\]

The product tail-biting trellis for this generator matrix is given in
Figure~\ref{fig:hamming-overlay}.
\begin{figure}[!ht]
	\centerline{\psfig{figure=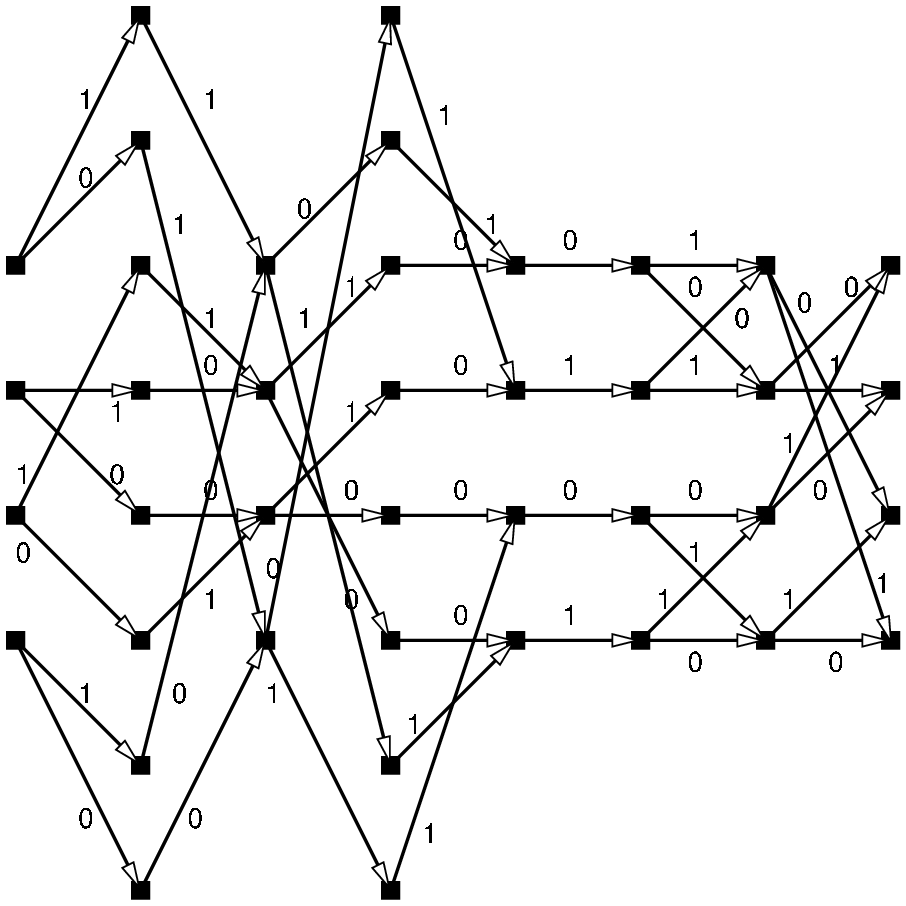,height=3.0in}}
	\caption{A product tail-biting trellis for the $\tup{7,4}$
	binary Hamming code.} 
	\label{fig:hamming-overlay}
\end{figure}
\end{example}

\begin{definition}
A subtrellis of a tailbiting trellis consists of a start node at time
index zero and all edges and nodes which can be traversed in any cycle
of the graph that begins and ends at this start node.
\end{definition}

\begin{figure}[!ht]
\vspace*{-0.3in}
\centerline{\psfig{figure=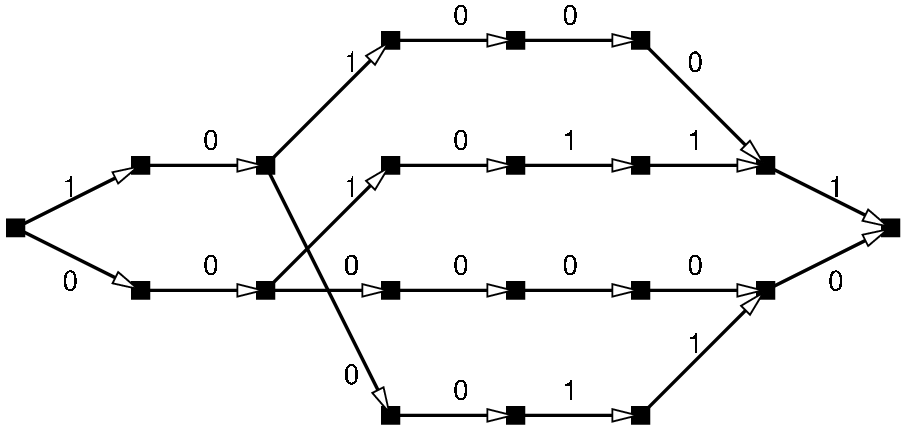,height=1.5in}}
\caption{Subtrellis $T_l = T_1$ of the tail-biting trellis for the (7,4) Hamming code in Figure~\ref{fig:hamming-overlay} for vectors of linear
  span}
\label{fig:subtrellis0}
\end{figure}

\begin{figure}[!ht]

\centerline{\psfig{figure=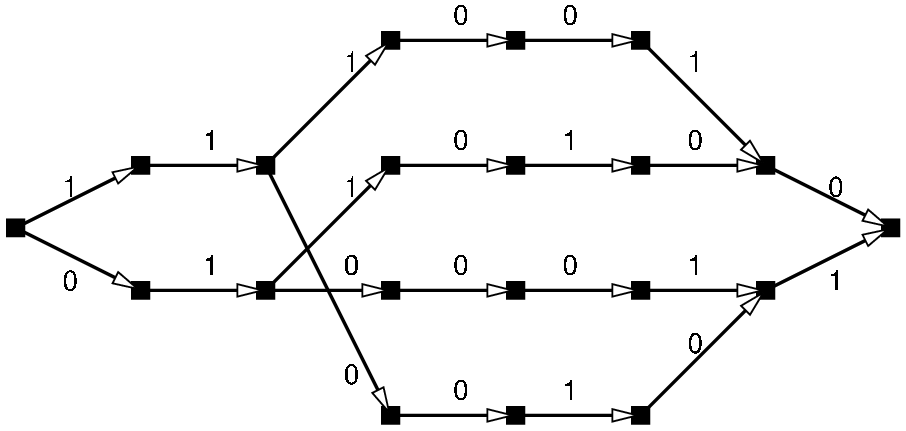,height=1.5in}}
\caption{Subtrellis $T_2$ corresponding to coset leader 0100011 with
  zero-run [2,5]}
\label{fig:subtrellis1}
\end{figure}

\begin{figure}[!ht]
\centerline{\psfig{figure=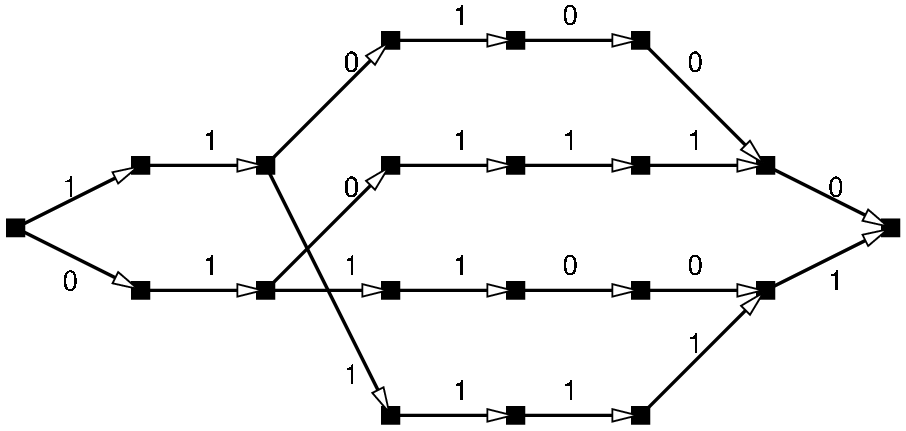,height=1.5in}}
\caption{Subtrellis $T_3$ corresponding to coset leader 0111001 with
  zero-run [4,6]}
\label{fig:subtrellis2}
\end{figure}

\begin{figure}[!ht]
\centerline{\psfig{figure=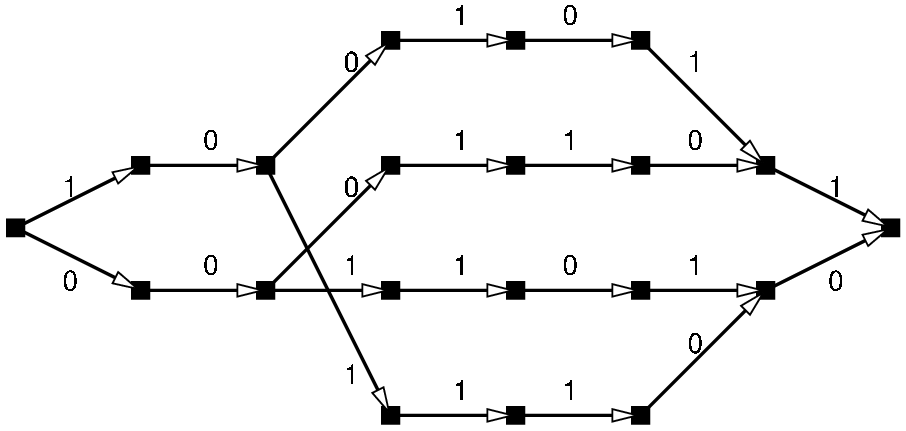,height=1.5in}}
\caption{Subtrellis $T_4$ corresponding to coset leader 0011010 with
  zero-run [4,5] }
\label{fig:subtrellis3}
\end{figure}

 Let $T_l$ denote the minimum conventional trellis for the code generated by $G_l$.
Clearly $T_l$ is a subtrellis of the tail-biting trellis.
If $l$ is the number of rows of $G$ with linear span and $c$ the
number of rows of circular span, the tail-biting trellis constructed
using the product construction will have $q^c$ start states. Each such
start state defines a subtrellis whose codewords form a coset of the
subcode corresponding to the subtrellis containing the all 0
codeword. The coset structure is well known and has been reported in
\cite{shao,tcs03,stacs,sb,solomon}. Each vector in the circular span can be
  considered  to be a coset leader . The set of zero runs,
 of the coset leaders  determines the structure of the tail-biting
  trellis in the following way. If a coset leader has zero run $[i,j]$
  then the subtrellis associated with that coset shares all states at time
  indices in the
  interval $[i,j]$ with the subtrellis corresponding to the subcode
  defined by vectors of linear span. Further, we recall, the {\em
    coset leader} shares
  all states in the interval $[i,j]$ with the states corresponding to
  the all-zero codeword.

 The  four
subtrellises of the tail-biting trellis of Figure
~\ref{fig:hamming-overlay} are shown in
Figures~\ref{fig:subtrellis0},~\ref{fig:subtrellis1},~\ref{fig:subtrellis2}and
~\ref{fig:subtrellis3} along with their associated coset leaders and
zero runs.

\begin{definition}

 If subtrellises $T_1 $ and $T_2$ share states from time
indices $i$ to $j$ then the interval $[i,j]$ is called the {\em
  merging} interval of $T_1$ and $T_2$.
\end{definition}
 It is easy to see that two subtrellises do not
share any states outside their merging interval.\\

 A tail-biting trellis is said to satisfy the {\em intersection
   property} if the intersection of all the zero runs of the members
 of $G_c$ is non-empty. The tail-biting trellis for the Hamming code
   given in Example ~\ref{ex:overlay2} satisfies the intersection
   property as the interval $[4,5]$ is contained in the intersection of
   all the zero runs of $G_c$.

\section{Decoding}
\label{algo}

The decoding algorithm proposed here is different from the sub-optimal algorithms
mentioned in Section~\ref{related}, that go round
and round the tail-biting trellis updating all the nodes of the trellis in every
round. It makes one round of the tail-biting trellis and subsequently
judiciously uses the information gathered to further update as few nodes as
it can before it closes in on the most likely codeword. Our
algorithm has two phases. In the first phase a Viterbi algorithm is
performed on the tail-biting trellis. This phase performs computations
at {\em every} node of the tail-biting trellis. In the second phase however,
{\em only one
  path} is tracked at a time, this being the most likely
path. The initial estimate of the most likely path is obtained from
the first phase.  This
path is present in some subtrellis and is followed until the 
algorithm decides that some other path (perhaps in another
subtrellis) looks more promising based on some metric. When such a
situation is encountered, computation on this path is
suspended and the more promising path is taken up. While this strategy
at first
glance looks
like the stack algorithm \cite{jelinek} for decoding convolutional
codes, it differs from it because it has the property that it {\em always}
delivers the optimal path as the metric used satisfies the property required by the
$A^*$ algorithm. (We will prove this property formally).
 
For purposes of decoding we use the unrolled version of the trellis
with start states $s_0,s_1\ldots s_l$ and final states $f_0,f_1 \ldots
f_l$ where $l$  is the number of subtrellises. An $(s_i,f_i)$ path is
a path from start vertex $s_i$ to final vertex $f_i$, and is
consequently  a codeword path in trellis $T_i$, whereas an $(s_i,f_j)$
path for $i \neq j$ is a non codeword path as it starts in subtrellis
$T_i$ and ends in subtrellis $T_j$. For purposes of our discussion we term the label sequence along such a path as a {\em semicodeword}.

Maximum-likelihood decoding for a tail-biting trellis is equivalent to
finding the codeword closest to the received sequence measured in
terms of a soft decision metric. Assume that the channel is modeled as
an additive white noise Gaussian(AWGN) channel and that antipodal
signaling is used for communication. A binary code digit 0 is mapped
into $\sqrt{E_s}$ and a 1 is mapped into $-\sqrt{E_s}$ where $E_s$ is
the signal energy per bit entering the channel. For a discrete
additive white Gaussian noise(AWGN) channel we have \[r_t = x_t +
n_t\] where $r_t$ is the received signal at time $t$, $x_t$ is the
transmitted signal and $n_t$ is the value of a white Gaussian noise random variable with variance $N_0/2$ where $N_0$ is the noise spectral density. Without loss of generality we can assume that $E_s = 1$. The signal-to-noise ratio or SNR is the quantity $E_s/N_0$.
The decoder uses the received vector $\vec{r}$ to determine which codeword was transmitted. It forms an estimate $\hat{\vec{x}}$ of the codeword $\vec{x}$ that was transmitted. A decoding error occurs if $\vec{x} \neq \hat{\vec{x}}$. The maximum likelihood decoding rule is to decode the received sequence $\vec{r}$ to codeword $\vec{x_m}$ whenever $p(\vec{r}/\vec{x_m}) \geq p(\vec{r}/\vec{x_l})$ for all $l \neq m$, where $p(\vec{r}/\vec{x_m})$ is the conditional probability of $\vec{r}$ given $\vec{x_m}$. Let $S(\vec{x})$ be the signal vector corresponding to the codeword $\vec{x}$.. 
If $d_E(S(\vec{x_m}),\vec{r})$ is the Euclidean distance between $S(\vec{x_m})$ and $\vec{r}$, then the maximum likelihood decoding rule for decoding binary linear block codes transmitted over the AWGN channel using antipodal signaling is to decode $\vec{r}$ into codeword $\vec{x_m}$ whenever $d_E(S(\vec{x_m}),\vec{r})\leq d_E(S(\vec{x_l}),\vec{r})$ for all $l\neq m$. 

 The decoding algorithm is thus cast as a shortest path problem in
 which each path is associated with a {\em metric}, and the problem is
 to find a codeword path with minimum metric. The
 $A^*$ algorithm is used to cut down the search space. It does so by using a node metric which is the
 sum of the length of the shortest path from the source to a node
 and an {\em underestimate} of the length of the shortest path from
 the node to the goal node to guide the search. As mentioned earlier,
 only one path is explored at a time and the algorithm derives it's
 advantage from the
 fact that if the estimates used are close to the actual values then
 the search space that yields the optimal path is greatly reduced. We
 give the algorithm below. The algorithm maintains two sets of
 vertices, $S$ and $\bar{S}$. The set $S$ is the set of {\em closed}
 nodes and represents nodes to which the shortest paths have been
 finalized. At any iteration, the set $\bar{S}$ is the set of
 candidate nodes the best of which will be closed in the succeeding
 iteration. These are called the {\em open} or {\em visited} nodes. An operation of
 {\em expanding} a node consists of the following three steps:\\
1. Getting all the immediate successors of the node.\\
2. Checking for each immediate successor if this successor has been
visited before.\\
3. If the successor has been visited then {\em updating} the minimum
cost path to the successor by taking the minimum of the cost of the previous path
and the cost of this one.
All the expanded nodes are put into the closed set and the visited nodes
are put into the open set. When the goal node is reached an optimal path
has been found.\\

 The following is a formal description  of
the algorithm. Line 1 performs the initialization of the sets and the
costs and paths. Line 3 selects the vertex to be expanded. Line 4 puts
the selected vertex into the closed set and deletes it from the open set. Line 5 detects if the
algorithm has completed; lines 6 through 9 perform an expansion of a
node. They update the cost of an
immediate successor as well as the best path to that successor and
mark the successor as visited by putting it into the open set.\\

\noindent
{\bf Algorithm $A^{*}$} \\
{\bf Input :} A trellis $T = (V, E, l)$ where $V$ is the set of 
vertices, $E$ is the set of edges and $l(u,v) \geq 0$ for edge $(u,v)$ 
in $E$, a source vertex $s$ and a destination vertex $f$, and an estimate $e(u,f)$ for the shortest path from $u$ to $f$ for each vertex $u \in V$. \\
{\bf Output :} The shortest path from $s$ to $f$. \\
/* $cost(u)$ is the cost of the current shortest path from $s$ to $u$ and $P(u)$
is a current shortest path from $s$ to $u$ */ \\
\emph{begin} \\
\hspace*{0.3in}{\bf 1.} $S \gets \emptyset, \quad \bar{S} \gets \lbrace s \rbrace, 
\quad cost(s) \gets 0, \quad P(u) \gets ( ), \forall u\in V, \quad cost(u) = +\infty , \forall u\neq s$; \\
\hspace*{0.3in}{\bf 2.}\emph{repeat} \\
\hspace*{0.6in}{\bf 3}. Let u be the vertex in $\bar{S}$ with minimum value of
$cost(u) + e(u,f)$. \\
\hspace*{0.6in}{\bf 4}.  $S \gets S \cup \lbrace u \rbrace; \quad \bar{S} \gets 
 \bar{S} \setminus \lbrace u \rbrace$; \\
\hspace*{0.6in}{\bf 5}. \emph{if} $u = f$ \emph{then} \emph{return} $P(f)$; \\
\hspace*{0.6in}{\bf 6}. \emph{for} each $(u,v) \in E$ \emph{do} \\
\hspace*{0.9in}{\bf 7}. \emph{if} $v \notin S$ \emph{then} \\
\hspace*{1.2in}{\bf 8}. \emph{begin} \\
\hspace*{1.5in}{\bf 9}. $cost(v) \gets \min(cost(u) + l(u,v), previous(cost(v)))$; \\
\hspace*{1.5in}{\bf 10.} \emph{if} $cost(v) \neq previous(cost(v))$ \emph{then} append $(u,v)$ to $P(u)$ to give $P(v)$; \\
\hspace*{1.5in}{\bf 11}. $(\bar{S}) \gets (\bar{S}) \cup \lbrace v \rbrace$; \\
\hspace*{1.2in}{\bf 12}. \emph{end} \\
\hspace*{0.3in}{\bf 13}. \emph{forever} \\
\emph{end} \\

The $A^*$ algorithm is guaranteed to output the shortest path if the following two conditions hold:
 Let $L_{T}(u,f)$ be the shortest path length from $u$ to $f$ in $T$.
Let $e(u,f)$ be any lower bound such that 
$e(u,f) \leq L_{T}(u,f)$, and such that $e(u,f)$ satisfies the following
inequality, i.e, for $u$ a predecessor of $v$, 
$l(u,v) + e(v,f) \geq e(u,f)$. If both the above conditions are 
satisfied, then the algorithm $A^{*}$, on termination, is guaranteed to
output a shortest path from $s$ to $f$.\\ 

The algorithm proposed here is a variant of the $A^*$ algorithm, which
at any given instant, is executing an $A^*$ algorithm on exactly one of the subtrellises, with perhaps suspended executions of the algorithm on a set of other subtrellises. The subtrellis on which the algorithm is curently executing, appears the best in its potential to deliver the minimal cost path. 
Since the algorithm is not straightforward, we first give
an informal explanation of how it works. The algorithm has two
phases. The first phase performs a Viterbi algorithm on the
tail-biting trellis and examines surviving paths, called {\em
  survivors} here, at all states of the tail-biting trellis.
The first phase is described below. Let $*e$ denote the initial vertex
of edge $e$. Let $e*$ denote the vertex entered via edge $e$.\\

\noindent
{\bf Algorithm First Phase}\\
{\bf Input:} An unrolled tail-biting trellis with start nodes
$s_1,s_2,\ldots s_l$, final nodes $f_1,f_2,\ldots f_l$ for the $l$ subtrellises,and an edge cost $c(e)$
associated with each edge $e$ of the tail-biting trellis.\\
{\bf Output:} The cost $cost(v)$ of a least cost path to each node $v$
from any start node.\\
\hspace*{0.5cm}\kwd{begin}\\
 \hspace*{1.0cm}\kwd{for} each node $v$ in the tail-biting trellis
 initialize $cost(v)=0$ ;\\
\hspace*{1.0cm}\kwd{for} $i$ = 1 to $n$ \kwd{do}\\
\hspace*{1.5cm}\kwd{for} each vertex $v$ at time index $i$ \kwd{do}\\
\hspace*{2.0cm} $cost(v)= min_{e:e*=v}\{cost(*e)+c(e)\}$\\
\hspace*{1.5cm}\kwd{for} $j$ = 1 to $l$ \kwd{do}\\
\hspace*{2.0cm}$metric(T_j)= cost(f_j);$\\
\hspace*{0.5cm}\kwd{end}

At the end of the first phase therefore we have a set of survivors at
final nodes $f_1,f_2,\ldots f_l$ some of which may not correspond to
codewords.
The costs of these paths are taken as initial estimates for the second phase.
We first informally describe the second phase below and then describe
a recursive version in more detail.\\

\noindent
{\bf Algorithm Second Phase}\\
{\bf Input:} The initial metrics $metric(T_i), i=1 \ldots l$ computed
in the first phase for the
$l$ subtrellises and the costs $cost(v)$ of the survivors at all
vertices $v$ of the tail-biting trellis.\\
{\bf Output:} The maximum likelihood path.\\
\hspace*{0.75cm}{\bf 1.} Sort the metrics $metric(T_i), i=1 \ldots l$ in
 increasing order; if the lowest metric is that of a codeword path
 then output that path as the ML path and return, else go to next
 step.\\
\hspace*{0.75cm}{\bf 2.} $low$ = cost of lowest codeword survivor if
there is one, otherwise, otherwise $low = \infty$. \\
\hspace*{0.75cm}{\bf 3.} If any of the metrics $metric(T_j)$ is
greater than $low$ then discard subtrellis $T_j$ from the set of
participants in the second phase.\\
\hspace*{0.75 cm}{\bf 4.} $Residual$-$trellises$ = set of all non-discarded
trellises with non-codeword survivors;\\
\hspace*{0.75cm}{\bf 5.} Create a set $\bar{S}$ of the initial
 vertices along with metrics, of all
 residual trellises, and let the start node $s$ of the $A^*$
 algorithm be the start node
 of the residual trellis with a minimum initial metric;\\
\hspace*{0.75cm}{\bf 6.} Execute lines 2 to 11 of Algorithm $A^*$
modifying statement in line 11 as 
{\bf if} $cost(v) < low$ {\bf then} $(\bar{S}) \gets (\bar{S}) \cup \lbrace v
\rbrace$ and statement $u=f$ in line 5 by $u\in \{f_1,f_2,\ldots f_l\}$\\
\hspace*{0.75cm}{\bf 7.} If the open set $(\bar{S})$ becomes empty before a final node is reached, then the codeword with cost $low$ is output as the decoder's estimate of the transmitted codeword.\\

The algorithm above is therefore different from the standard $A^*$
algorithm in the following ways:
\begin{enumerate}
\item It may  switch from one subtrellis  to
another depending on which subtrellis the node with minimum metric is
located in.
\item  Each shared node in a subtrellis is regarded as a
distinct node for purposes of the algorithm. Thus there will be as
many distinct copies of a given node of the tail-biting trellis as there are
residual subtrellises sharing that node.
\item Before adding an element
to the open set, we check to see that its metric is less than that of
the best codeword survivor stored in $low$. In the traditional algorithm there is no such check.
\item If the open set $\bar{S}$ becomes empty before a final node is reached then the codeword with cost $low$ is output.
\end{enumerate}

We
 need to define the estimate $e(u,f)$ in line 3 of Algorithm
$A^*$. Recall that this has to be an {\em underestimate} of the path
length from node $u$ to the final node if the ML path is to be output.
The estimate we use for node $v$ in subtrellis $T_j$ is the difference
between the {\em initial} metric for trellis $T_j$ computed in the first phase
 and the cost of the {\em survivor} at node $v$ in the first
phase. We will prove later that this is {\em indeed} an underestimate and
therefore guarantees that the ML path is output on termination. We
implement the open set $\bar{S}$ as a {\em heap}~\cite{aho}. This ensures that
the minimum element can be retrieved in {\em constant} time and that whenever
an element is {\em inserted} into the heap, restructuring it in order to
preserve the property of
constant time access to the minimum element, has complexity logarithmic
in the size of the heap.\\

\noindent
We now describe the second phase of the algorithm more formally beginning with the notation used.\\
1. Variable $e(s_i,f_i)$ is the estimate obtained for the shortest path from the start to the final node in subtrellis $T_i$ in the first phase.\\
2. Variable $e(v,f_i)$ is the estimate for the shortest path from node $v$ to node $f_i$ in subtrellis $T_i$ which is computed when an update occurs at node $v$. This is the difference between the initial estimate at $s_i$ in trellis $T_i$, and the cost of the survivor at node $v$ in the first phase.\\
3. Variable $h$ is a pointer to a structure representing a node in the trellis; $h.state$ is the state, $h.trellis$ indicates which trellis that state belongs to; $h.metric$ stores the current metric which is the sum of the length of the path from the start node in trellis $h.trellis$  to $h.node$ and the estimate of the path length from $h.node$ to the final node in that trellis. \\
4. Variable $succ$ is a pointer to the successor of a node; $succ.state$ and $succ.metric$ have meanings that can be deduced from 3 above.\\
5. Variable $index$ refers to the time index and takes on values from 0 to $n-1$ where $n$ is the length of the code.\\
6. Variable $trellisnumber$ is a unique number associated with a subtrellis.\\
7. Function $InsertHeap$ inserts a node into the heap; function $DeleteMin$ extracts the node with minimum value of metric from the heap.\\
8.  Function $IsEmpty$ returns a boolean value which is true if the heap is empty and false otherwise.\\
9. Variable  $node.cost$ represents the actual cost of the path from the start state of a subtrellis that ends at the node $node$. Variable $node.cost1$ represents the cost of the survivor in the first phase at that node.\\
10. Variable $metric$ is the updated metric at a successor of a node in a trellis using function $Update$, which is called when that node is closed using $Expand$. \\
11. Variable $P(state)$ is the sequence of nodes representing the winning path at the state $state$.\\
12. Variable $low$ is the cost of the lowest cost $(s_i,f_i)$ path in the first phase.\\
13. Variable $flag$ is used to detect whether the winning path is the one identified in the first or second phase. It is initialized to 0. If the heap becomes empty without reaching a final node in the second phase then the lowest cost $(s_i,f_i)$ path is output as flag remains 0. Else the path that first reaches a final node in the second phase is  the winning path.

\noindent
\kwd{function}$ Second\_Phase$\\
/* Begin with $r$ residual trellises whose metrics have been sorted in
increasing order, and with variable $low$ which stores the metric of
the best codeword survivor*/ \\
\kwd{begin}\\
/* First create a heap $H$ with these $r$ metrics; each element of the heap is a record containing the trellis number, the node, the time index, and the metric*/\\
\hspace*{1.0cm} \kwd{for} $i= 1 $ \kwd{to} $r$ \kwd{do}\\
\hspace*{1.5cm} $InsertHeap(H, i, startVertex(T_i),0, e(s_i,f_i))$\\
\hspace*{1.0cm} \kwd{endfor}\\
\hspace*{1.5cm} $flag= 0$;\\
\hspace*{1.0cm} \kwd{while} $IsEmpty(H)=$ \kwd{false} \kwd{and} $flag = 0$ \kwd{do}\\
\hspace*{1.5cm}$ h := \ DeleteMin(H)$\\
\hspace*{1.5cm}$ S := S\cup h.node$ /*Add $h.node$ to the set of closed nodes*/\\
\hspace*{1.5cm}$Expand(h.trellisNo,h.state,h.timeindex,h.metric)$ /* Expand $h.node$*/ \\
\hspace*{1.0cm}\kwd{endwhile}\\
\hspace*{1.0cm} \kwd{if} $flag=0$ \kwd{then} output the codeword with
metric $low$; \kwd{return}\\
\kwd{end}\\

\noindent
\kwd{function}$Expand(trellisnumber, state, index, metric)$\\
\hspace*{0.5cm}1.\kwd{begin}\\
\hspace*{1.0cm}2. \kwd{if} $index = n-1$ \kwd{then} $flag = 1$; output $P(state)$; \kwd{return}\\
\hspace*{1.0cm}3. \kwd{else}\\
\hspace*{1.5cm}4. \kwd{for} each successor $succ$ of $state$ \kwd{do}\\
\hspace*{2.0cm}5. $Update(trellisnumber,state,succ.state,succ.metric,index)$\\
\hspace*{2.0cm}6. \kwd{if} $succ.metric\leq metric$ \kwd{then}$S:=S\cup \{succ.state\};$\\
\hspace*{2.0cm}7.  $Expand(trellisnumber,succ.state,index,succ.metric)$ \\
\hspace*{2.0cm}8. \kwd{else}\\
\hspace*{2.5cm}9.   \kwd{if} $succ.metric < low$\\
\hspace*{2.5cm}10.   \kwd{then}$InsertHeap(H,trellisnumber,succ.state,index,succ.metric)$\\
\hspace*{2.5cm}11.\kwd{endif}\\
\hspace*{2.0cm}12.\kwd{endif}\\
\hspace*{1.5cm}13.\kwd{endfor}\\
\hspace*{1.0cm}14.\kwd{endif}\\
\hspace*{0.5cm}15. \kwd{end}\\

\noindent
\kwd{function}$Update(i,node1, node2, metric,timeindex)$;\\
\hspace*{0.5cm}\kwd{begin}\\
\hspace*{1.0cm}$timeindex:= timeindex+1$\\
\hspace*{1.0cm}$newcost := node1.cost + edgecost(node1,node2)$\\
\hspace*{1.0cm}\kwd{if} $newcost \leq node2.cost$ \kwd{then}\\
\hspace*{2.0cm}$P(node2):= (P(node1),node2)$ /* update the current shortest path to $node2$*/\\ 
\hspace*{2.0cm}$node2.cost:= newcost$ /* update the cost of the current shortest path to node 2*/ \\
\hspace*{2.0cm}$metric:= node2.cost+ e(s_i,f_i)- node2.cost1$/* update the metric at $node2$; $node2.cost1$ is the cost of the survivor in the first phase*/\\
\hspace*{1.0cm}\kwd{endif}\\
\hspace*{0.5cm}\kwd{end}

\section{Analysis of the Decoding Algorithm}
\label{analysis}

We first prove that on termination the algorithm always outputs the optimal path

\begin{lemma}
Each survivor at a node $u$ has a cost which is a lower bound on the cost of the least cost
path from $s_{j}$ to $u$ in an $(s_{j},f_{j})$ path passing through $u$.
\end{lemma}

\begin{proof}
Assume that $u$ is an arbitrary node on an $(s_{j},f_{j})$ path and that
path $P$ is the survivor at $u$ in the first phase. There are two cases.
Either $P$ is a path from $s_{j}$ to $u$ or $P$ is a path from $s_{i}$ 
to $u$, $j \neq i$. If the latter is the case, then the cost of $P$ is
less than the cost of the path from $s_{j}$ to $u$; hence the cost of
the survivor at $u$ is a lower bound on the cost of the least cost path
from $s_{j}$ to $u$.
\end{proof}

\begin{lemma}
\label{properties}
The quantity $e(u,f_j)$  defined in the algorithm satisfies the following two
properties :
\begin{enumerate}
\item $e(u,f_{j}) \leq L_{T_j}(u,f_{j})$
\item $l(u,v) + e(v,f_j) \geq e(u,f_j)$ where $(u,v)$ is an edge. 
\end{enumerate}
\end{lemma}

\begin{proof}
\begin{enumerate}
\item $e(u,f_j)= cost(survivor(f_j))-cost(survivor(u))$\\
Also $cost(survivor(f_j))\leq cost(survivor(u)) +L_{T_j}(u,f_j)$, from which the result follows.

\item  To prove: $\quad l(u,v) \ + \ e(v,f_{j}) \ \geq \ e(u,f_{j})$ \\
\vspace*{0.1in} 
\hspace*{1.0in} LHS $\ = \ l(u,v) \ + \ e(v,f_{j})$ \\
\hspace*{1.4in}$ = \ l(u,v) \ + \ e(s_{j},f_{j}) \ - \ cost(survivor(v))$\\ 
\vspace*{0.1in} 
If survivor at $v$ is survivor at $u$ concatenated with edge $(u,v)$, 
then\\
\vspace*{0.1in} 
\hspace*{0.7in} LHS $\ = \ l(u,v) \ + \ e(s_{j},f_{j}) \ - \ 
	cost(survivor(u)) \ - \ l(u,v)$ \\
\hspace*{1.1in}$ = \ e(u,f_{j})$\\ 
\vspace*{0.1in} 
On the other hand  if survivor at $v$ is not a continuation of the 
survivor at $u$,
\vspace*{0.1in} 
\begin{center}
$cost(survivor(v)) \ < \ cost(survivor(u)) \ + \ l(u,v)$\\
\vspace*{0.1in} 
$cost(survivor(v)) \ - \ l(u,v) \ < \ cost(survivor(u))$\\
\vspace*{0.1in} 
or, $\quad e(s_{j},f_{j}) \ - \ cost(survivor(v)) \ + \ l(u,v)
	\ > \ e(s_{j},f_{j}) \ - \ cost(survivor(u))$\\

\vspace*{0.1in} 
or, $\quad e(v,f_{j}) \ + \ l(u,v) \ > \ e(u,f_{j})$\\
\vspace*{0.1in} 
Therefore, $\quad l(u,v) \ + \ e(v,f_{j}) \ \geq \ e(u,f_{j})$
\end{center}
\end{enumerate}
\end{proof}

Lemma~\ref{properties}, and the fact that all estimates on trellises
on which execution is suspended are underestimates, assures us that
{\em if the final node is reached in any subtrellis then this is indeed the shortest path in the tail-biting trellis or in other words the ML codeword.}

We first make a few observations about the algorithm. During any point in the second phase, the algorithm  is exploring some path in a candidate subtrellis called the {\em current} trellis even though it may do so in discontinuous steps. This path is called the  {\em current path} in that  subtrellis. The metric which it uses to decide whether to continue on the current path on the current trellis, say $T_i$, or forsake it in favour of another path either in the current trellis or on another candidate trellis is initially $e(s_i,f_i)$. We have the following lemma specifying how the metric changes along the path.
\begin{lemma}
\label{update}
 During the second phase, if the current path updates a node $v$ using function $Update$, where the survivor in the first phase was not in the current subtrellis then the metric becomes $e(s_i,f_i)+\Delta(i,v)$ where $\Delta(i,v)$ is the difference between the cost of the least cost path ending at $v$ in the current trellis and the survivor at $v$ during the first pass.
\end{lemma}
\begin{proof}
 We know that
 \begin{equation}
  cost(s_i,v)= cost(s_i,u)+edgecost(u,v)
  \end{equation}
   and
  \begin{equation}  
    e(v,f_i)= e(s_i,f_i)- cost(survivor(v))
   \end{equation}  
     The metric is just the sum of the two lefthand sides of the previous two equations. Thus if the survivor is the current path then
     \begin{equation}
      cost(survivor(v))= cost(s_i,u) + edgecost(u,v)
      \end{equation}
      and the lemma follows. If the survivor is not the current path then the metric is increased by the difference between the length of the current path up to $v$ and the survivor at $v$.
\end{proof}
\begin{definition}
A critical node on a path in a subtrellis is one at which the metric for a subtrellis
 reaches its final value(i.e. the actual cost of the path).
\end{definition}
\begin{lemma}
\label{lem:critical}
During the second phase, once a critical node is closed in a subtrellis, the algorithm
goes on to reach the final node in that subtrellis without switching trellises, and outputs
an ML path.
\end{lemma}
\begin{proof}
The critical node was closed because it had the minimum metric. 
The metric represents 
the {\em actual} cost of the path at a critical node.  This is
no greater than the metrics of all other visited nodes which are {\em underestimates} of the costs of all other paths. Thus once a critical node is closed, the metric does not change along the continuation of this winning path to the final node. Therefore line 6 of function $Expand$ is always true at some successor andno trellis switching takes place.
\end{proof}

The following properties hold for the metric. Let $m_i(N)$ denote the metric in subtrellis $i$ at node $N$:
\begin{lemma}
\label{predecessor}
Let an $(s_k,f_i)$ path  be the winner at $f_i$ in the first phase and let it win over an $(s_i,f_i)$ path at node $A$. Then $m_i(A)= m_i(f_i)$ and $m_i(B)<m_i(f_i)$ for any proper predecessor $B$ of $A$.
\end{lemma}
\begin{proof}
Since the $(s_k,f_i)$ path was the overall winner at $f_i$ its length
will be the metric at the start node of trellis $T_i$ and by
Lemma~\ref{update}, the metric on the path in $T_i$ will rise by the
appropriate amounts $\Delta_j$ at each node $j$ where the path was
overtaken by a path from some other subtrellis. When it reaches node
$A$, which is a critical node, the metric will reach its final value, namely $m_i(f_i)$. Since $B$ is a predecessor of $A$ and the metric {\em rises} at $A$, $m_i(B)<m_i(f_i)$.
\end{proof}

For each shortest path in a subtrellis $i$, the nodes where it was
overtaken by paths originating at the start nodes of other subtrellises
in the first phase, are the nodes where its metric will rise during
the second phase. These nodes are called {\em rising points}. Thus the node at the final rising point in a subtrellis is the critical node.

\begin{lemma}
 \label{share}
Let subtrellises $T_i$ and $T_j$ share a node $N$ and between them, let $T_i$ be the first to close the node in the second phase. Then $m_i(N)\leq m_j(N)$.
\end{lemma}
\begin{proof}
Since $T_i$ is the first to close the node it closes it either before $T_j$ was first opened or after. If the former was the case, then $m_i(N)\leq m_j(s_j)\leq m_j(N)$. If the latter was the case the least current metric of $T_j$ is greater than the metric $m_i(N)$ of $T_i$ from which the result follows as the metric can only increase.
\end{proof}
\begin{lemma}
\label{segment}
For nodes $A$ and $B$ let $(A,B)$ be a path segment in the merging interval of $T_i$ and $T_j$ and let $m_i(A)\leq m_j(A)$. Then $m_i(B)\leq m_j(B)$.
\end{lemma}
\begin{proof}
Since at $A$, $m_i(A)\leq m_j(A)$ and thereafter all updates to the metrics in trellises $T_i$ and $T_j$ until node $B$ is reached will be identical as the survivors at those node in the first phase will be the same for both trellises $T_i$ and $T_j$, $m_i(B)\leq m_j(B)$.
\end{proof}

 We next show that any path from an arbitrary start node to any final node represents a vector in a vector space. For the sake of simplicity we restrict our arguments to binary codes.

\begin{lemma}
\label{vectorspace}
The set of all labels from an arbitrary start node to any final node is a vector space.
\end{lemma}
\begin{proof}
Assume that each of the $c$ vectors in the submatrix $G_c$ of the
generator matrix is of the form  $v_i = [\vec{h_i},\vec{0},\vec{t_i}]$
where $v_i$ has circular span $[j,k]$,
where $\vec{h_i}$ stands for the sequence of symbols from the first, up to and
including the $k^{th}$ symbol
 and is called the {\em head}, and $\vec{t_i}$ stands for the
sequence of symbols from positions $j$ to $n-1$ and is called the {\em
  tail}; $\vec{0}$ represents the run  of zero symbols in between the
head and the tail, spanning the appropriate
number of codeword indices. (This run may be empty if $j = k+1$.)  Let $\{v_1,v_2 \ldots v_c\}$ be the vectors of $G_c$. Then the matrix $G_s$ defined as
$G_s = \left[\begin{array}{c} G_l \\ \hline G_c' \end{array}\right]$, where $G_c'$ consists of $2c$ rows of the form $[\vec{h_i},\vec{0}],[\vec{0},\vec{t_i}], 1\leq i \leq c$, (where the number of zeroes in $\vec{0}$ makes up a total of $n$ elements for the row)
generates the set of labels of all paths from any start node to any final node. This set has $2^{l+2c}$ elements.
This can be verified from the product construction. The set of elements of this vector space consists of {\em semicodewords} and codewords.
Each semicodeword is the label of an $(s_i,f_j)$ path $i \neq j$.
\end{proof}

\begin{example}
The matrix $G_s$ corrresponding to the matrix $G_{KV}$ for the Hamming
(7,4) code of Example~\ref{ex:overlay2} is displayed below.\\
\[
G_{s} = \left[\begin{array}{ccccccc}
1 & 0 & 0 & 0 & 1 & 1 & 0 \\
0 & 0 & 1 & 0 & 1 & 1 & 1 \\
\hline
0 & 1 & 0 & 0 & 0 & 0 & 0 \\
0 & 0 & 0 & 0 & 0 & 1 & 1 \\
0 & 1 & 1 & 1 & 0 & 0 & 0\\
0 & 0 & 0 & 0 & 0 & 0 & 1\\
\end{array}
\right]
\]

It can be observed that the semicodeword 1100110 formed by adding rows
1 and 3 of $G_s$
traces a path from start vertex $s_2$ to final vertex $f_1$ in the tail-biting trellis of
Figure ~\ref{fig:hamming-overlay}.
\end{example}
\begin{lemma}
\label{not-close}
The algorithm will not close any node whose metric exceeds the cost of the ML path.
\end{lemma}
\begin{proof}
The lemma follows from lines 6 and 7 of function $Expand$ and the
observation that calling function $Expand$ on a node is equivalent to
closing the node. The test ensures that only nodes with metric value
less than the current metric are closed. Since the current metric is a
lower bound on the cost of the ML path the lemma follows.
\end{proof}

We use a result of Tendolkar and Hartmann~\cite{tendolkar} stated below.

\begin{lemma}
\label{tendolkar}
Let $H$ be the parity check matrix of the code and let a codeword $\vec{x}$ be transmitted as a signal vector $S(\vec{x})$. Let the binary quantization of the received vector $\vec{r}= r_1,r_2,\ldots r_n$ be denoted by $\vec{y}$. Let $\vec{r'}= (|r_1|,|r_2|,\ldots |r_n|)$ and $S=\vec{y}H^T$. Then ML decoding is achieved by decoding a received vector $\vec{r}$ into the codeword $\vec{y}+\vec{e}$ where $\vec{e}$ is a binary vector that satisfies $\vec{s}= \vec{e}H^T$ and has the property that if $\vec{e'}$ is any other binary vector such that $\vec{s}= \vec{e'}H^T$ then $\vec{e}.\vec{r'} < \vec{e'}.r'$ where $.$ is the inner product.
\end{lemma}
A direct consequence of Lemma~\ref{tendolkar} is the following result.
\begin{lemma}
\label{ml-codeword}
If the all-zero codeword is the  ML codeword for an error pattern $\vec{e}$ then\begin{equation}
\label{equ:ml}
 \vec{e}.r' < (\vec{c} +\vec{e}).r'
\end{equation}
 for any non-zero codeword $\vec{c}$.
\end{lemma}

Since the space explored by the algorithm, namely the space of
semicodewords and codewords is a vector space, we can analyse the
algorithm assuming that the ML codeword is the all 0 codeword.
\begin{lemma}
\label{semicodewords}
Assume the all 0 codeword is the ML codeword. Let $\vec{e}$ be the binary quantization of the received vector.
For the error pattern $\vec{e}$  the second phase of the decoding algorithm will close the start nodes of only those subtrellises whose initial metric corresponds to a semicodeword $C_s$ satisfying
\begin{equation}
\label{equ:weights}  
(C_s +\vec{e}) .r' < \vec{e}.r'
\end{equation}

\end{lemma}
\begin{proof}
We first note that at the start of the second phase the metrics at the
start nodes of all residual subtrellises correspond  to the costs of vectors
in the vector space of codewords and semicodewords, i.e. the vector space defined by the generator matrix $G_s$.
From Lemma~\ref{vectorspace} we have $(C_s +\vec{e})H_s^T=
\vec{e}.H_s^T$ where $H_s$ is the parity check matrix corresponding to
the matrix $G_s$. From Lemma~\ref{tendolkar} maximum likelihood
decoding on the set of semicodewords will initially choose $C_s$, a
semicodeword, which satisfies the inequality of the Lemma and the algorithm will close the start node of the subtrellis with that initial metric. As the algorithm proceeds with updating metrics it may close start nodes of other subtrellises. However by Lemma~\ref{not-close} it will never close the start node of any trellis $T_j$ whose initial metric exceeds that of the ML codeword, which implies that the all-0 codeword is more likely than the semicodeword survivor in $T_j$, thus implying Equation~\ref{equ:weights}.
\end{proof}

The properties of the algorithm proved in this section will be used to
explain the good performance of the approximate algorithms described
in the following section.

\section{An Approximate Algorithm}
\label{approx}
Recall that each shared node is treated as a distinct node in the
second phase of the algorithm. We now  propose an approximate variant of the exact algorithm which
closes a shared node {\em at most once} in the second phase. We term this
algorithm $Approx1$.\\

Assume  we replace line 5 of function $Expand$ by\\
 \kwd{if} $succ.state\notin S$ \kwd{then} $Update(trellisnumber,state,succ.state,succ.metric,index)$ \kwd{else} continue\\

What this ensures is that each shared node is closed {\em at most
 once}, that is, by at most one subtrellis, in
the second phase. Therefore the total number of Viterbi updates in the
first phase and expansions in the second phases is at most $2V$ where
 $V$ is the number of states in the tail-biting trellis. Since a node
 is closed by at most one subtrellis, it is conceivable that a shared node
 that is on the ML path is closed by a subtrellis that does not
 contain the ML codeword. In such a case the result produced will not
 be the ML codeword. We now analyse the conditions under which this
 happens. The symbols are the same as those defined for Lemma~\ref{semicodewords}.

The following theorem gives the conditions under which the approximate algorithm produces a non-ML output. Recall that the intersection property requires that the intersection of all the zero runs of vectors in $G_c'$ be non-empty.

\begin{theorem}
\label{explanation}
 If the tail-biting trellis satisfies the intersection property, the
 approximate algorithm produces a non-ML output for error patterns
 $\vec{e}$ satisfying equation ~\ref{equ:weights} whenever $C_s$ is a
 semicodeword which is formed as a linear combination of rows of $G_s$
 that contain at least one non-zero multiple of a vector from $G_l$.
\end{theorem}
\begin{proof}
Let us assume that the all-zero codeword is the ML codeword but that
 it is not the output of the approximate algorithm
 $Approx1$. Therefore some trellis say $T_i$ must close a node $N$ on
 the all 0 path (so that $T_0$ never gets to close it, as only one closure is allowed, and therefore cannot output the all 0 path). Clearly node $N$ must be in the merging interval of $T_0$ and $T_i$. Since $T_i$ is a residual trellis(otherwise it would have not participated in the second phase), let the survivor at $f_i$ in the first phase be an $(s_k,f_i)$ path that overtakes the $(s_i,N,f_i)$ path at node $A$, in other words, $A$ is the critical node for trellis $T_i$.
 
{\bf Case 1.} Suppose node $A$ is a predecessor of node $N$. By Lemma
 ~\ref{predecessor},  $m_i(A) = m_i(f_i)$, and since $A$ is a critical
 node, by Lemma ~\ref{lem:critical}, $T_i$ would have gone on to win in the exact algorithm and therefore the all-zero codeword could not have been the ML codeword giving a contradiction.

{\bf Case 2.} Suppose node $A$ is a successor of $N$ {\em within} the merging interval of $T_i$ and
 $T_0$. By Lemma~\ref{predecessor} $m_i(A)= m_i(f_i)$. Since $\vec{0}$
 is the ML codeword $m_i(f_i)>m_0(f_0)$ implying that
 $m_i(A)>m_0(f_0)$. Since subtrellis $T_i$ closed node $N$, by
 Lemma~\ref{share},  $m_i(N)\leq m_0(N)$. By the property of the
 metric $m_0(N)< m_0(f_0)$ implying that $m_i(N)<m_0(f_0)$. Since $A$
 is in the merging interval of $T_0$ and $T_i$  by Lemma~\ref{segment}
 $m_i(A)\leq m_0(A)\leq m_0(f_0)$ giving a contradiction. Therefore we
 conclude that if subtrellis $T_i$ closes $N$ and $A$ is a successor
 of $N$,  then $A$ cannot be in the merging interval of $T_i$ and $T_0$.\\

We thus conclude that $A$ is beyond the merging interval of $T_0$ and
 $T_i$, and hence the $(s_k,A,f_i)$ path does not touch the all-zero path. Since the intersection property is satisfied, any path which is a linear combination of vectors of $G_c'$ alone must have at least one node on the all-zero path. Hence the semicodeword corresponding to the  $(s_k,A,f_i)$ path cannot be formed as a linear combination of rows only in $G_c'$ and therefore it is formed as a linear combination of vectors with at least one member of $G_l$.
\end{proof}

Theorem~\ref{explanation} and Lemmas~\ref{ml-codeword} and ~\ref{semicodewords} provide an
explanation of the experimental observation that decoding differences
between the exact and the approximate algorithm are infrequent, so
much so, that the bit error rate curves are practically
indistinguishable. Lemma~\ref{semicodewords} tells us that in order
for a subtrellis to be opened it must contain a semicodeword
satisfying equation~\ref{equ:weights}, being the most likely
semicodeword among the possible candidates. Theorem~\ref{explanation}
establishes the condition that if a node on the all-zero path is
closed by some trellis $T_i$ other than $T_0$ when the all-zero
codeword was transmitted, then the initial metric of $T_i$ must be
that of a semicodeword of pretty high weight (because it is a linear
combination of vectors which contain at least one vector in $G_l$). Further, the error $\vec{e}$ which caused the cost of this high weight semicodeword to drop significantly enough to satisfy Equation~\ref{equ:weights}, should not cause the weight of any non-zero {\em codeword} to drop by an amount enough to violate Equation~\ref{equ:ml}. Since semi-codewords share prefixes and suffixes with codewords, such events may be quite infrequent.\\

One could get an even better approximation by allowing a node to be
closed at most twice. We have experimented with this and observe that
the bit error rate for this approximation is indistinguishable from
that of the exact algorithm at all values of signal to noise
ratio for all the three codes on which we have run the simulations. The significance of this is that the time complexity can be
explicitly bounded by the complexity of at most three computations for each
node of the tail-biting trellis, one update in the first Viterbi
decoding phase and at most
two expansions in the second phase.
\subsection{Complexity Analysis}
We now estimate the time complexity of the approximate algorithm.
The following bound on the complexity of the Viterbi algorithm is well known\cite{mceliece}.
\begin{lemma}
\label{viterbi}
The complexity of the first phase of the decoding algorithm is $O(E)$
where $E$ is the number of edges in the tail-biting trellis.
\end{lemma}

The next lemma is a statement of a well known result on heap data structures\cite{aho}.
\begin{lemma}
\label{heap}
Each insertion into the heap has complexity $O(\log H)$ where $H$
is the number of elements in the heap.
\end{lemma}
\begin{theorem}
\label{approx1}
The algorithm $Approx1$ has complexity bounded by $O(E \log V)$ where
$V$ is the number of states in the tail-biting trellis.
\end{theorem}
\begin{proof}
The number of vertices that are updated is at most $2V$ as each vertex is expanded at most once in the
second phase. Each time a vertex is expanded it results in
computations on every edge leaving it and at most a constant
number of elements being visited and inserted into the heap
$\bar{S}$,(as this number is bounded by the field size
assumed to be a constant). The complexity of each insertion 
phase is $\log H$ where $H$ is the size of the heap. Since this size is
proportional to $V$ the complexity of the second phase is $O(E \log
V)$. The sorting operation at the end of the first phase has complexity
$O(V_0\log V_0)$ where $V_0$ is the number of states at time index
$0$. The complexity is dominated by the $O(E \log V)$ term and hence
the theorem.
\end{proof}

To reduce the overheads, the heap is implemented as $m$ separate heaps
if there are $m$ residual trellises, with a separate heap of pointers,
each element of which  points to the root of a distinct subtrellis heap. The individual
heap sizes are small in practice and the algorithm is practically
linear in the size of the trellis. In the next section we present
results from profiling the program which bear out the claim that the
overheads of heap operations are negligible.

An argument similar to that in Theorem~\ref{approx1} estalishes the
complexity of algorithm $Approx2$ as $O(E \log V)$.
We next look at the space complexity of the algorithm.\\
\begin{lemma}
\label{lem:space-complexity}
The space requirement for algorithm $Approx1$ is $O(V_0\times V)$ bits.
\end{lemma}
\begin{proof}
The algorithm requires $O(V)$ space to store the estimates at each
state in the first phase. The additional space required  to store the
heap is also $O(V)$ as each expanded node can
put at most all its successors on the heap. The bit vectors that store
trellis membership are of size $V_0$ where $V_0$ is the number of
start nodes of the tail-biting trellis. The space requirements for the
bit vectors is therfore $V_0\times V$ bits. The space requirements for
storing the current cost at each node is $O(V)$. This follows from the
fact that each shared node is
closed at most once. This means that at most one copy of a shared node
updates its succesors. 
This in turn means that each successor has at most one update along
each of its incoming edges. Since the number of incoming edges is a
constant
which is at most the size of the field, a constant number of costs are
associated with each node in the tail-biting trellis from which
the result follows.
\end{proof}

\section{Simulations}
\label{simulations}
\begin{figure}[ht]
\centerline{\psfig{figure=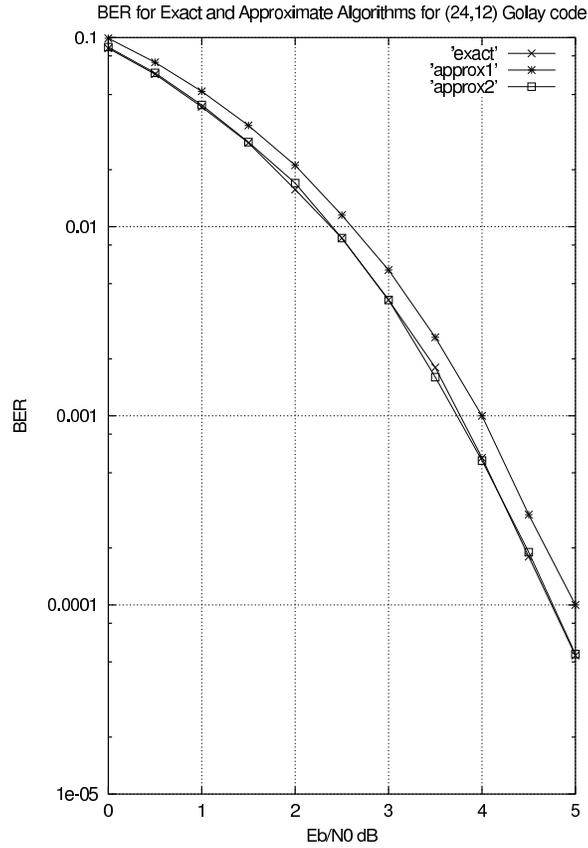,width=3in}}
\caption{BER for the Exact and Approximate Algorithms for the (24,12) Extended Binary Golay Code}
\label{fig:golay-ber}
\end{figure}

\begin{figure}[!ht]
\centerline{\psfig{figure=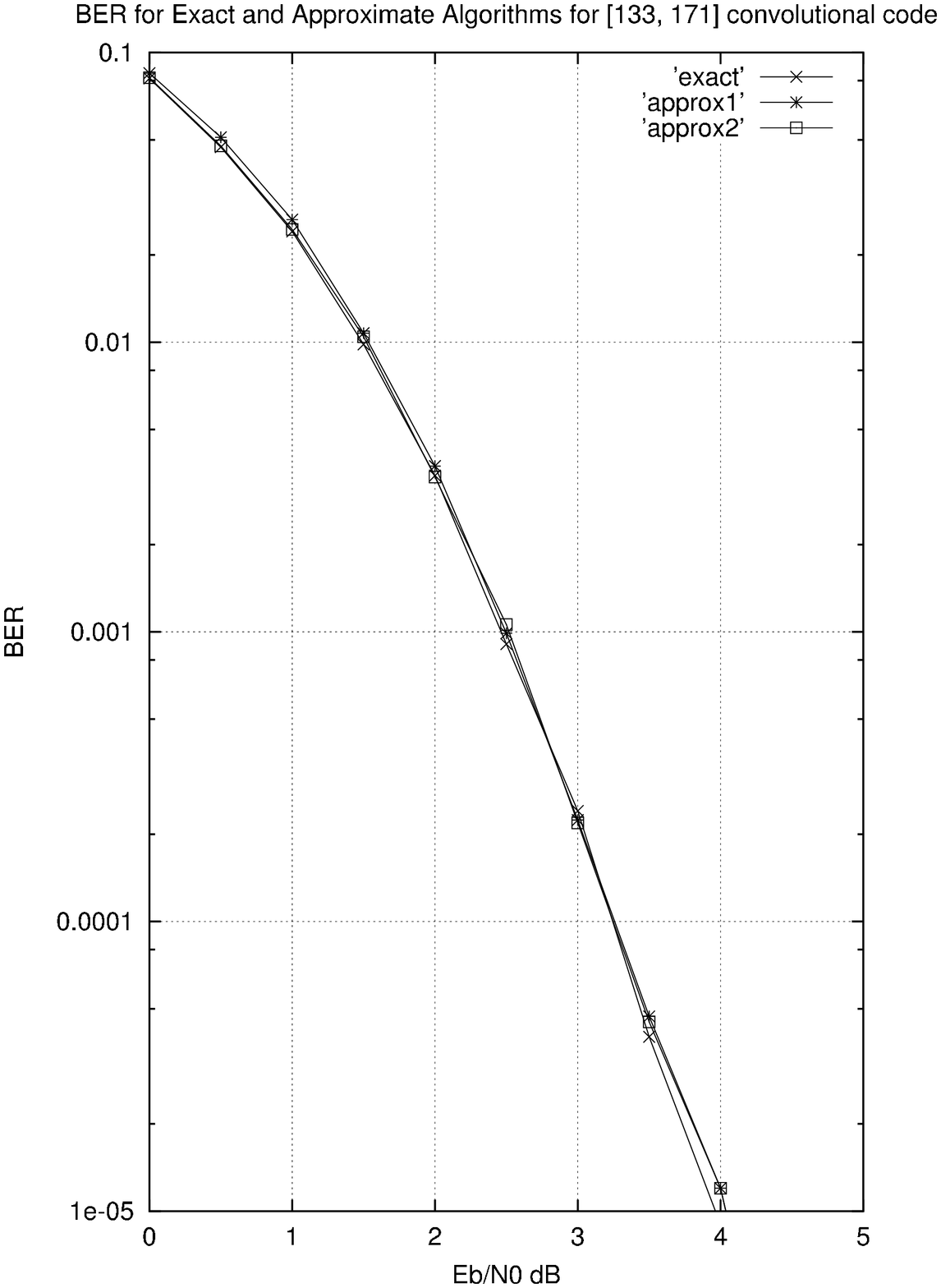,width=3in}}
	\caption{Bit Error Rates for the Exact and Approximate Algorithms for the rate 1/2 (133,171) Convolutional Code with circle length 48}
	\label{fig:conv1-ber}
\end{figure}

\begin{figure}[!ht]
\centerline{\psfig{figure=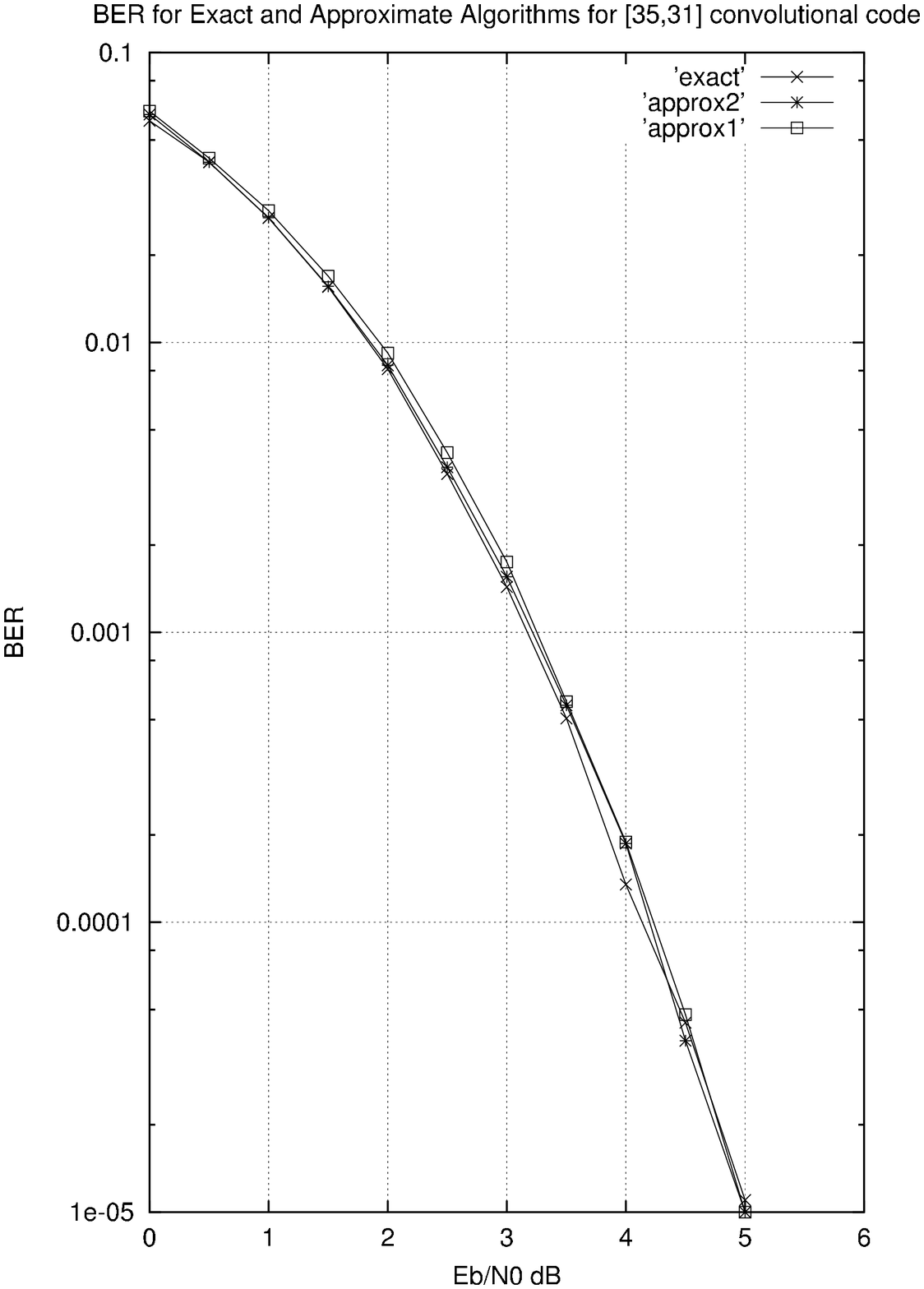,width=3in}}
	\caption{Bit Error Rates for the Exact and Approximate Algorithms for the rate 1/2 (35,31) Convolutional Code with circle length 20}
	\label{fig:conv2-ber}
\end{figure}

We have coded the exact and approximate algorithms and show the
 results of simulations on  minimal tail-biting trellises for the 16
 state tail-biting trellis~\cite{cald} for the extended (24,12) Golay
 code on an AWGN channel with antipodal signaling, and tail-biting
 trellises for two rate 1/2 convolutional codes with memory 6, circle
 size 48 (which is the same as the (554,744) convolutional code
 experimented with in~\cite{map}, and memory 4, circle size 20 (which is
 the same as the (72,62) convolutional code used in~\cite{cav}
 respectively. We show the variation of both, the average as well as
 the maximum number of node computations (counting Viterbi updates in
 the first phase and expansions in the second phase) with the
 signal to noise ratio for our exact algorithm, and compare this with
 the number of Viterbi updates needed for the brute force
 approach. Note that this number is indicative of the {\em time complexity}
 of the algorithm.   The results are encouraging and are displayed in
 Tables~\ref{tab:expansions1},~\ref{tab:expansions2} and
 ~\ref{tab:expansions3} respectively for the Golay code and the two
 convolutional codes. {\em On the average, the number of updates to
 get the exact ML result requires fewer than two computations at each node
 of the tail-biting trellis at all values of signal to noise
 ratio, one in the first pass and one in the second}. The maximum number of node computations for the algorithm
 $Approx1$ is obviously bounded by twice the number of nodes in the
 tail-biting trellis. We also display the bit error-rate performance of the
 approximate algorithms closing nodes at most once for the first
 approximation $Approx1$, and at most twice for the second
 approximation, $Approx2$ in Figures ~\ref{fig:golay-ber},~\ref{fig:conv1-ber},~\ref{fig:conv2-ber} and  and find that there is virtually no difference in the bit error rates for the second approximation and the exact ML algorithm. {\em Thus we get virtually ML performance for an explicit linearly bounded update complexity at all values of signal to noise ratio}.



\begin{table}[!ht]
\begin{center}
\begin{tabular} {|c|c|c|c|}
\hline 

{\small \bf SNR}&   {\small \bf Maximum Heap Size}& {\small \bf Maximum Node Computations}& {\small \bf Average Node Computations} \\ 	
\hline
0.0&	285&	602	&245.2\\
\hline
0.5&	294	&688	&235.3\\
\hline
1.0&	311	&709	&225.7\\
\hline
1.5&	273	&637	&217.7\\
\hline
2.0&	271	&580	&210.6\\
\hline
2.5&	256	&576	&204.8\\
\hline
3.0&	289	&643	&200.1\\
\hline
3.5&	242	&557	&197.2\\
\hline
4.0&	192	&480	&195.1\\
\hline
4.5&	152	&423	&193.8\\
\hline 
5.0&	135	&396	&193.0\\
\hline
\end{tabular}
\caption {Runtime statistics for the Exact algorithm for the $(24,14)$ Extended Binary Golay Code. A brute force algorithm would typically perfrom $1744 $ updates. The tail-biting trellis has $192$ states.
\label{tab:expansions1}}
 
\end{center}
\end{table}

\begin{table}[!ht]
\begin{center}
\begin {tabular} {|c|c|c|c|}
\hline
{\small \bf SNR}&   {\small \bf Maximum Heap Size}& {\small \bf Maximum Node Computations}& {\small \bf Average Node Computations} \\ 	
\hline
0.0&	13064&	22311  &4414.1 \\
\hline
0.5&	15698&	24958  	&4051.4 \\
\hline
1.0&	13161&	20369	&3738.5 \\
\hline
1.5&	12926&	18981	&3487.9 \\
\hline
2.0&	9948&	16162	&3330.0 \\
\hline
2.5&	7492&	11700	&3233.5 \\
\hline
3.0&	5743&	11175	&3175.0 \\
\hline
3.5&	3354&	7163	&3138.2 \\
\hline
4.0&	2781&	6447	&3115.0 \\
\hline
4.5&	1526&	5104	&3099.5 \\
\hline
5.0&	1059&	4693	&3088.2 \\
\hline
\end{tabular}
\caption{Runtime statistics for the Exact algorithm for the rate $1/2$ $[133,171]$
convolutional code with circle length $48$. A brute force algorithm would
typically perform $159552$ updates. The tail-biting trellis has $3072$ states.
\label{tab:expansions2}}
\end{center}
\end{table}

\begin{table}[!ht]
\begin{center}
\begin{tabular} {|c|c|c|c|}
\hline 
{\small \bf SNR}&   {\small \bf Maximum Heap Size}& {\small \bf Maximum Node Computations}& {\small \bf Average 
Node Computations} \\ 	
\hline
0.0&	701	&1437	&426.9 \\
\hline
0.5&	784	&1447	&405.4 \\
\hline
1.0&	824	&1554	&384.9\\
\hline
1.5&	749	&1426	&367.6\\
\hline
2.0&	623	&1214	&353.5\\
\hline
2.5&	563	&1179	&342.7\\
\hline
3.0&	578	&1162	&334.6\\
\hline
3.5&	503	&984	&329.5\\
\hline
4.0&	412	&918	&326.2\\
\hline
4.5&	292	&718	&323.7\\
\hline
5.0&	241	&660	&322.3\\
\hline
\end{tabular}
\caption{Runtime statistics for the Exact algorithm for the rate $1/2$ $[35,31]$ convolutional code with circle length $20$. A brute force algorithm would
typically perform $4368$ updates. The tail-biting trellis has $320$ states.
\label{tab:expansions3}}
\end{center}
\end{table}
 
\section{Discussion and Conclusions}
\label{conclusion}
We have proposed an exact algorithm for ML decoding on tail-biting
trellises and also experimented on two approximate variants. The
average time complexity of the exact algorithm is seen to be quite
low. The
 approximate variants perform as well as the exact one in terms
of the bit error rate at an explicitly bounded update complexity
equivalent to two, or sometimes three rounds on the tail-biting trellis. The algorithm does not suffer from the effects of
limit cycles or pseudocodewords which current iterative algorithms are
subject to. Profiling measurements carried out on the program are
displayed in Table~\ref{tab:profile}. The execution time was averaged
over 10,000 runs of the decoder.  The percentage of execution
time taken up by each of the five major operations in the decoding
process, namely, the initializations of all the arrays, the first pass, the sorting operation at the end of
the first pass, the second pass, and the heap operations is displayed. It can be observed that heap
operations incur an overhead of only 11 \% of the program running time
at 0 dB and are negligible for higher values of signal to noise
ratios.

\begin{table}
\begin{center}
\begin{tabular} {|c|c|c|c|c|c|}
\hline 
 \ &{\bf Percentage}& {\bf of}& {\bf Execution}& {\bf Time}& \  \\
\hline
{\bf SNR}&   {\bf Initializations}& {\bf Phase 1}& {\bf
  Sorting} &{\bf Phase 2} &{\bf Heap operations} \\ 	
&&&&& \\
\hline
0.0&    14.75\%   &34.09\%	&5.88\%	&31.38\%	&11.12\%\\

\hline
1.0&	8.34\% 	&46.95\%	&8.34\% 	&24.94\%	&6.61\%\\

\hline
2.0&	6.07\%	&58.84\%	&10.20\%	&17.62\%	&2.74\%\\

\hline
3.0&	3.21\% 	&69.02\%	&10.79\%	&12.68\%	&0.92\%\\

\hline
4.0&	1.13\%	&74.23\%	&11.49\%	&9.62\%	&0.23\%\\

\hline 
5.0&	0.52\% 	&75.70\%	&11.59\%	&8.425\%	&0.09\%\\
\hline 
\end{tabular}
\label{tab:profile}
\caption{Profiling statistics for the two phase decoder for Algorithm
  Approx1}
\end{center}
\end{table}

The results of simulations on the extended (24,12) Golay code, a rate 1/2, memory 6 convolutional code with a circle size of 48(which is the same as the (554,744) convolutional code used for experiments in \cite{map} and a rate 1/2 memory 4 convolutional code with a circle size of 20(which is the same as the (72,62) rate 1/2 convolutional used for experimentation in \cite{cav}) have been reported. It is seen that the second approximate variant has a bit error rate which is indistinguishable from that of the exact algorithm for all values of signal to noise ratio.\\

{ \bf Acknowledgement} The authors gratefully acknowledge discussions
with Aditya Nori. They would also like to thank the anonymous referees
for their comments which greatly improved the presentation of the paper.\\



\begin{thebibliography}{99}
\bibitem{aguado}L.E. Aguado and P. G. Farrell, On hybrid stack
  decoding algorithms for block codes, {\em IEEE
  Trans. Inform. Theory}, January 1998, pp 398-409.
\bibitem{aho}A. Aho, J.E. Hopcroft and J.D. Ullman, {\em Data
  Structures and Algorithms}, Addison-Wesley, Reading, MA. 1983. 
\bibitem{aji}S. Aji, G. Horn, R. McEliece and M. Xu, Iterative Min-Sum Decoding of Tail-Biting Codes, {\em Proceedings of Information Theory Workshop}, Killarney, Ireland, June 22-26, pp. 68-69.

\bibitem{cav}J.B. Anderson and S.M. Hladik, An Optimal Circular Viterbi Decoder for the Bounded Distance Criterion, {\em IEEE Transactions on Communications}, {\bf 50}(11), November 2002.
\bibitem{map}J.B. Anderson and S.M. Hladik, Tail-biting MAP Decoders,{\em IEEE Journal in Selected Areas in Communication}, {\bf16}(2), February 1998. 
\bibitem{cald} A.R. Calderbank, G.D. Forney,Jr., and A. Vardy,
Minimal Tail-Biting Trellises: The Golay Code and More, 
{\em IEEE Trans. Inform. Theory}, {\bf 45}(5), July 1999, pp. 1435-1455.

\bibitem{sundberg} R.V. Cox and C.V. Sundberg, An Efficient Adaptive Circular Viterbi Algorithm for Decoding Generalized Tailbiting Convolutional Codes, {\em IEEE Transactions on Vehicular Technology} {\bf 43}(1), February 1994, pp 57-68.
\bibitem{stacs}
Kaustubh Deshmukh, Priti Shankar, Amitava Dasgupta, and B.~Sundar Rajan.
 On the many faces of block codes.
In {\em Symposium on Theoretical Aspects of Computer Science(STACS)},
  pages 53--64, 2000.
\bibitem{ekroot}L. Ekroot and S. Dolinar, $A^*$ decoding of block
  codes, {\em IEEE Trans. Commun.} {\bf 44} (9), September 1996, pp
  1052-1056.
\bibitem{han1} Y.S. Han, C.R.P. Hartmann and C.-C. Chen, Efficient
  priority-first search maximum-likelihood soft-decision decoding of
  linear block codes, {\em IEEE Trans. Inform. Theory} {\bf 39} (5)
  September 1993, pp 1514-1523.
\bibitem{han2} Y.S. Han, C.R.P. Hartmann, and K.G. Mehrotra, Decoding
  linear block codes using a priority first search: Performance
  analysis and suboptimal version, {\em IEEE Trans. Inform. Theory}
  {\bf 44}(7), November 1998, pp 3091-3096.
\bibitem{han3} Y.S. Han, A new treatment of priority-first search
  maximum-likelihood soft-decision decoding of linear block codes,
  {\em IEEE Trans. Inform. Theory} {\bf 44}(7) November 1998, pp
  3091-3096.
\bibitem{han4}Y.S. Han, P.-N. Chen and H. -B. Wu, A maximum-likelihood
  soft-decision sequential decoding algorithm for binary convolutional
  codes, {\em IEEE Trans. Commun.} {\bf 50} (2) , February 2002, pp
  173-178.
\bibitem{han5} Y.S. Han, C.R.P. Hartmann and K.G. Mehrotra, Decoding
  Linear Block Codes Using a Priority-First Search: Performance
  Analysis and Suboptimal Version, {\em IEEE Trans. Inform. Theory}
  {\bf 44}(3), May 1988, pp 1233-1246. 
\bibitem{jelinek}F. Jelinek, A fast sequential decoding algorithm
  using a stack, {\em IBM J. Res. Devel} {\bf 13}, 1969, pp 675-685.
\bibitem{kv1} R. Koetter and A. Vardy, On the theory of linear trellises,
{\em Information, Coding and Mathematics} (M. Blaum, Editor), Boston:Kluwer,
May 2002.

\bibitem{kv2} R. Koetter and A. Vardy,  The Structure of Tail-Biting Trellises:
Minimality and Basic Principles, IEEE Trans. 
Inform. Theory, September 2003, pp. 2081-2105.

\bibitem{ksch95}F.R. Kschischang and V. Sorokine, On the trellis structure of
block codes, {\em IEEE Trans. Inform. Theory}, {\bf 41}(6), Nov 1995, 
pp. 1924-1937.
\bibitem{ksch98} F.R. Kschischang and V. Sorokine, A sequential decoder
  for linear block codes with a variable bias term metric, {\em IEEE
  Trans. Inform. Theory} January 1998, pp 410-411.

\bibitem{ku}B.D. Kudryashov, Decoding of block codes obtained from convolutional codes, {\em Problemy Peredachi Informatsii},{\bf 26}(2), pp 18-26, April-June 1990(in Russian). English Translation, Plenum Publishing Corporation, Oct 1990.

\bibitem{mceliece}R.J. McEliece, On the BCJR Trellis for Linear Block
  Codes, {\em IEEE Trans. Inform. Theory}, {\bf 42} (4), July 1996, pp 1072-1092.
\bibitem{ma}J.H.Ma and J.K.Wolf, On tail-biting convolutional codes, {\em IEEE Trans. Commun.},{\bf 34},February 1986, pp 104-111.
\bibitem{nil}N.J. Nilsson, Principles of Artificial Intelligence,
  Tioga Publishing Co., Palo Alto, CA, 1980.
\bibitem{tcs03}P. Shankar, A. Dasgupta, K. Deshmukh and B.S. Rajan, On Viewing Block Codes as Finite Automata, Theoretical Computer Science, {\bf 290}(2003) 1775-1797.

\bibitem{sh2} P. Shankar, P.N.A. Kumar, K. Sasidharan and B.S. Rajan, ML decoding of block codes on their tail-biting trellises, in {\em Proc. 2001 IEEE Int. Symposium on Information Theory}, IEEE Press, 2001, pp. 291.
\bibitem{shih} C.-C.Shih, C.R. Wulff, C.R.P. Hartmann and C.K. Mohan,
  Efficient heuristic search algorithms for soft-decision decoding of
  linear block codes, {\em IEEE Trans. Inform. Theory}, November 1998,
  pp 3023-3038.


\bibitem{sb}Yaron Shany and Yair Be'ery, Linear Tail-Biting Trellises, the Square-Root Bound, and Applications for Reed-Muller Codes, {\em IEEE Trans. Inform. Theory,} {\bf 46}(4), July 2000, pp 1514-1523.

\bibitem{shu}R.Y.Shao, Shu Lin and M.P.C. Fossorier, Two decoding algorithms for tail-biting codes,  {\em IEEE Trans. Commun.},{\bf 51}(10), October 2003, pp 1658-1665.

\bibitem{shao}Shu Lin and R.Y.Shao, General Structure and construction of Tail Biting Trellises for Linear Block Codes, in {\em Proc. 2000 Int. Symp. Inform. Theory}, Sorrento, Italy, pp 117.

\bibitem{solomon}G.Solomon and H.C.A. van Tilborg, A connection between block and convolutional codes, {\em SIAM J. Appl. Math.}, {\bf 37}, October 1979, pp 358-369.
\bibitem{tanner} M. Tanner, A recursive approach to low complexity codes, {\em IEEE Trans. Inform. Theory} {\bf 27}, pp 533-547, 1981.
\bibitem{tendolkar}N.N. Tendolkar and C.R.P. Hartmann, Generalization of Chase Algorithms for Soft Decision Decoding of Binary Linear Codes, {\em IEEE Trans. Inform.Theory,}{\bf 30}(5), September 1984,pp 714-721.


\bibitem{wang}Q.Wang and V.K.Bhargava, An efficient maximum-likelihood decoding algorithm for generalized tail-biting codes including quasi-cyclic codes, {\em IEEE Trans. Commun.}, {\bf 37}, August 1989, pp 875-879.


\bibitem{chep}K.S.Zigangirov and V.V. Chepyzhov, Study of decoding tail-biting convolutional codes, in {\em Proc. Swedish-Soviet Workshop on Information Theory}(Gotland, Sweden, Aug. 1989), pp 52-55.



\end{thebibliography}
\end{document}